\documentclass[11pt,twoside]{article}

\usepackage{asp2006}
\usepackage{epsf}
\usepackage{psfig}
\usepackage{lscape}

\begin{document}
\title{Adaptive Optics Assisted 3D spectroscopy observations for black hole mass measurements}  
\author{Guia Pastorini}   
\affil{Dipartimento di Astronomia e Scienza dello Spazio, Largo E.Fermi 5, 50125, Firenze, Italia}   

\begin{abstract} 
The very high spatial resolution provided by Adaptive Optics assisted spectroscopic observations at 8m-class telescopes (e.g. with SINFONI at the VLT) will allow to greatly increase the number of direct black hole (BH) mass measurements which is currently very small. This is a fundamental step to investigate the tight link between galaxy evolution and BH growth, revealed by the existing scaling relations between $M_{BH}$ and galaxy structural parameters. I present preliminary results from SINFONI K-band spectroscopic observations of a sample of 5 objects with $M_{BH}$ measurements obtained with the Reverberation Mapping (RM) technique. This technique is the starting point to derive the so-called virial $M_{BH}$ estimates, currently the only way to measure $M_{BH}$ at high redshift. Our goal is to assess the reliability of RM by measuring $M_{BH}$ with both gas and stellar kinematical methods and to investigate whether active galaxies follow the same $M_{BH}$-galaxy correlations as normal ones.
\end{abstract}

\section{\label{introduzione}Introduction and Description of the program}   
It is now widely accepted that the mechanism powering Active Galactic Nuclei (AGN) is accretion of matter onto a BH (Salpeter 1964), with masses in the $10^{6}-10^{10}M_{\odot}$ range (Ferrarese \& Ford 2005).  This paradigm, combined with the cosmological evolution of AGNs, implies that most, if not all, local galaxies should host a BH in their nuclei (e.g. Marconi \emph{et al.} 2004). BH are detected in about $40$ galaxies, mostly early-types (E/S0; e.g. Ferrarese \& Ford 2005). The BH mass ($M_{BH}$) is closely related to the stellar velocity dispersion ($\sigma_{*}$; Ferrarese \& Merritt 2000, Gebhardt \emph{et al.} 2000) and to the host bulge mass or luminosity ($L_{sph}$, Kormendy \& Richstone 1995, Marconi \& Hunt 2003). These correlations reveal a tight link between the growth of central BH and the formation and evolution of the host galaxy. 
The direct methods adopted to measure $M_{BH}$ in the nearby universe use gas or stellar kinematics to gather information on the gravitational potential in the nuclear region of the galaxy.
The stellar kinematical method has the advantage that stars are present in all galactic nuclei and their motion is always gravitational. The drawback is that it requires relatively long observation times in order to obtain high quality observations and that stellar dynamical models are very complex, potentially plagued by indeterminacy (Valluri \emph{et al.} 2004, Cretton \& Emsellem 2004).
Conversely, the gas kinematical method is relatively simple, it requires relatively short observation times, even if not all galaxy nuclei present detectable emission lines. Another important drawback is that non-circular or non-gravitational motions might invalidate this method. 
Since the observed correlations are based on BH masses obtained with different methods, it is important to check whether these methods provide consistent and robust results. This check so far has been performed only for the Milky Way (see Genzel \& Eckart 2000 and Genzel 2006) and Centaurus A (see Marconi \emph{et al.} 2006 and Silge \emph{et al.} 2005), where the two independent estimates are in excellent agreement.
All methods used to directly detect BHs and measure their masses require high spatial resolution in order to resolve the BH sphere of influence, i.e. the region where the gravitational influence of the BH dominates over that of the host galaxy. Its radius is traditionally defined as $r_{BH}=GM_{BH}/\sigma^{2}$ (where $G$ is the gravitational constant and $\sigma$ is the stellar velocity dispersion) and is usually $< 1\arcsec$ (e.g. Pastorini \emph{et al.} 2007). For a very long time the highest spatial resolution was provided by the \emph{Hubble Space Telescope} (HST) whose angular resolution is approximately $\delta\theta = 0.07\arcsec$ (Full Width at Half Maximum, FWHM, of the Point Spread Function, PSF) at $\lambda \sim 6500\AA$.
Notwithstanding this resolution results insufficient to investigate galaxies beyond the local universe and one has to rely on to indirect methods to estimate black hole masses at cosmological distances. In type 1 AGNs, under the assumption that the Broad Line Region (BLR) is virialized, the so-called \emph{virial} black hole mass can be computed as $M_{BH}=f \sigma^{2}R_{BLR}G^{-1}$, where $G$ is the gravitational constant, $R_{BLR}$ the size of the BLR, $\sigma$ the emission-line width and $f$ is a structural factor which takes into account the geometry of the BLR. 
$R_{BLR}$ can be computed from the time delay between the variation of the continuum emission and the variation of broad emission line. This delay is in fact interpreted as the time needed to the light to travel trough the BLR (see Peterson 2006 for a review). 
Recently correlations between $R_{BLR}$ and other physical quantities like the continuum emission at $5100 \AA$ ($R_{BLR} \propto L_{5100}^{0.7}$) or the luminosity  of the $H_{\beta}$ ($R_{BLR}\propto L(H\beta)^{0.63 \pm 0.06}$, see Kaspi 2006 for a review) have been discovered and provide a much easer way to estimate $R_{BLR}$ than the time consuming RM technique. These virial masses can be estimated at very high redshift.It is crucial to asses the reliability of RM estimates beyond what has been done with the comparison with $M_{BH}$-galaxy scaling relations. An important improvement in this field of research has been apported by Adaptive Optics (AO) assisted observations on very large telescope e.g. at the VLT which, with its 8 meter of diameter, offers a spatial resolution in K band of $0.06 \arcsec$ (see also Davies \emph{et al.} 2006). 
We have undertaken a spectroscopic survey of Seyfert 1 galaxies and quasars using SINFONI at the VLT. 
Our sample is composed of 5 AGN with very low redshifts ($z < 0.16$) with avalaible $M_{BH}$ Reverberation Mapping estimates.
Our aim is to use the high spatial resolution to apply both direct methods $M_{BH}$ measurements to our sample and to compare our estimates with the Reverberation Mapping ones. 
Here I present preliminary results obtained for two objects of our sample: 3C120 and NGC 4593.
3C120 (z = 0.033), classified as a FRI, is characterized by a one-sided superluminal jet (Seielstad \emph{et al.} 1979), extending from the core on subparsec scales to about 100 kpc, with a complex strucure (see Axon \emph{et al.} 1989 and Walker \emph{et al} 2001). 
NGC 4593 (z = 0.009) is a well-known Seyfert 1 galaxy (see Lewis \emph{et al.} 1978 or MacAlpine \emph{et al.} 1979) with a very bright nuclear point source which completely swamps the central brightness ditribution of the galaxy.
The $M_{BH}s$, estimated with Reverberation Mapping technique are $M_{BH}=5.02 \times 10^{7}M_{\odot}$  for 3C120 (see Peterson 2006) and $M_{BH}=5.2 \times 10^{6} M_{\odot}$ for NGC 4593 (see Denney \emph{et. al} 2006). 

\section{\label{analisi} Observations and Preliminary results}

We obtained VLT/SINFONI K-band spectra for both galaxies with and without AO (the pixel scales were respectively 0.025 and 0.25 arcsec). The spectra were analyzed using the \emph{spred} pipeline (Eisenhauer \emph{et al.} 2003).
The tracers of gas kinematics are the molecular hydrogen $H2$ $\lambda$ $21218$  $\AA$, the $Br\gamma$ $\lambda$ $21660$ $\AA$ and the coronal line $[SiVI]$ $\lambda$ $19635$ $\AA.$ 
The seeing limited spectra ($\delta\theta \sim 0.2 ''$) were used to constrain the stellar mass 
contribution to the gravitational potential under whose effects the gas is moving; conversely the diffraction limited ones have been used to constrain $M_{BH}$.
\subsubsection{\label{4593}Preliminary Results for NGC 4593} 

Kinematical parameters were obtained by fitting the emission lines in each row of the continuum-subctracted two
dimensional spectra and modelling the emission lines with 
a Gauss-Hermite serie (Cappellari \& Emsellem 2004).Following the technique proposed by van der Marel \& van den Bosch (1998) we estimated, from the data without AO, a stellar M/L of $M/L < 0.1 M_{\odot}/L_{\odot,K}$, consistent with a stellar population younger than $1 Gyr$. The molecolar hydrogen ($\lambda = 21218 \AA$) velocity map (Fig.\ref{fig:sil}) observed in this galaxy shows clearly the presence of a gas rotating disk. This hypothesis is supported by the fact that the center of rotation, the putative position of the black hole, is coincident with the peak of both surface brightness profile and velocity dispersion. A preliminary analysis allows us to give only an upper limit for the $M_{BH}$, which results $M_{BH}< 10^{8}M_{\odot}$, in agreement with the Reverberation Mapping estimate.

\subsubsection*{\label{3c120}Preliminary Results for 3C120}
   
\begin{figure}[!ht]
\plottwo{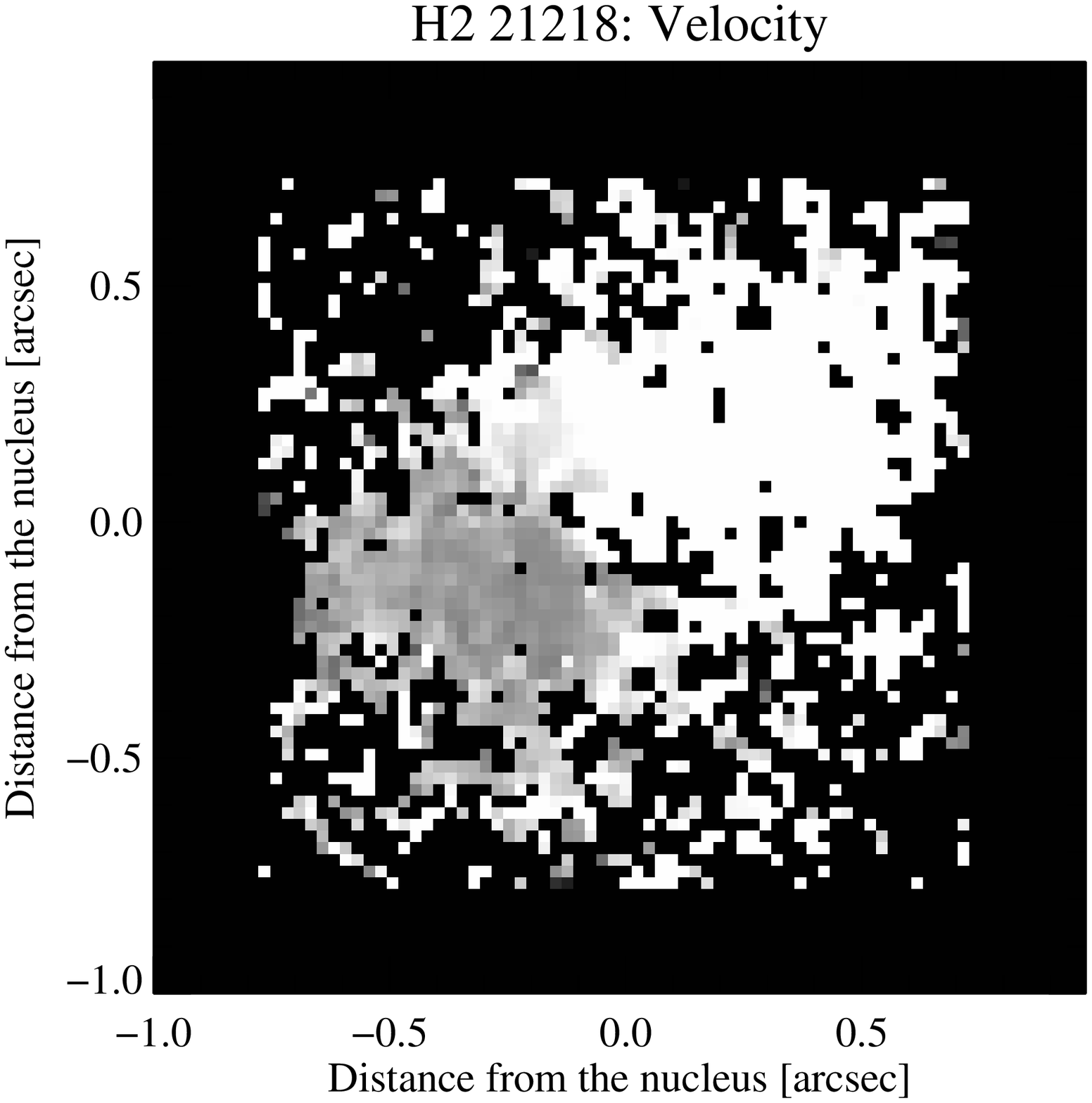}{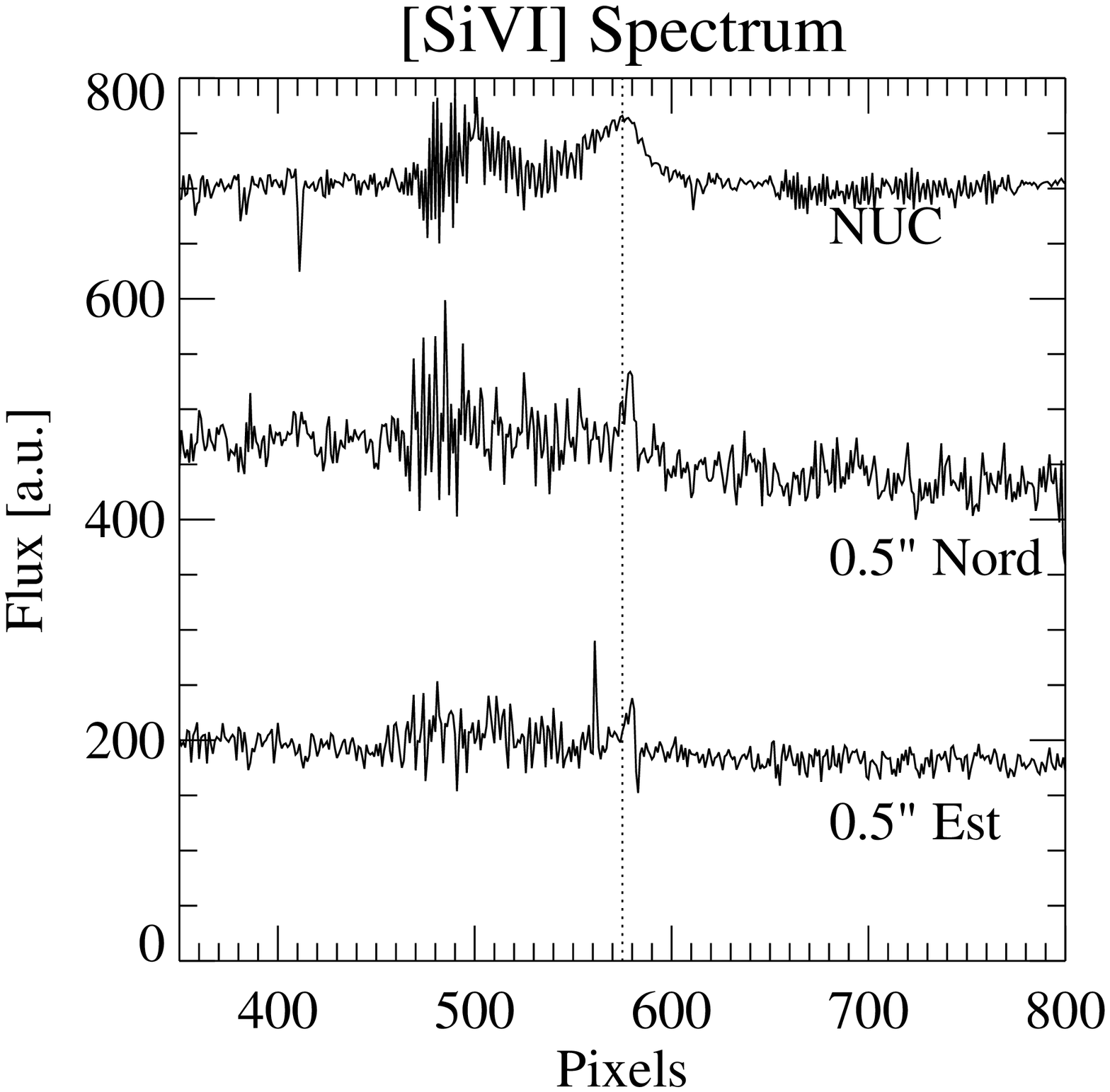}
\caption{\label{fig:sil}\emph{Right}: NGC 4593. Velocity map obtained from the molecular hydrogen. The presence of a rotating disk is identified by gas approaching us (darker gray) and gas moving away from us (lighter gray).\emph{Left}: 3C120. Comparison between the [SiVI] spectrum extracted from the nuclear region (top) and two external regions. Away from the nucleus, the emission line becomes systematically narrower and redshifted with respect to the nucleus.}
\end{figure}

In the case of 3C120, we focus our attention on the [SiVI] coronal line, which is characterized by a peculiar kinematics.
The velocity map does not show any indication of rotation. On the contrary, the velocity at the location of the AGN is lower than that from  the more distant regions. A more accurate analysis shows a blueshift of the nuclear emission with respect to the more extended one (see Fig.\ref{fig:sil}).
A possibile explanation is that the [SiVI] emission is made by two component. A nuclear one, unresolved in our observations, characterized by a FWHM of $\sim 1100 km/s$, coming from an unresolved region with size
 $r<50 pc$ placed in an intermediate region between the BLR and the NLR. As already found in coronal lines (e.g. Marconi \emph{et al.} 1996,Rodriguez-Ardila \emph{et al.} 2002), this component is blueshifted with respect to the systemic emission, indicating that it is composed by material moving in our direction, maybe caused by an outflow of material evaporating from the AGN torus. This ipothesis may be supported by the presence of a superluminal jet (see Axon \emph{et al.} 2004 for a review)
The second fainter component, is extended at galactic scales ($r \sim 300 pc$) and is coming from the NLR, as indicated by its FWHM ($\sim 400 km/s$). This extended component is moving at the systemic velocity of the galaxy (so we observed it as red-shifted); it likely partecipates to the galaxy rotation but any velocity gradients would be hidden by the more intense nuclear component.

\acknowledgements 
I thank F. Eisenhauer for providing very important suggestions and helps regardind the SINFONI data reduction.

\end{document}